\documentclass[pdflatex,sn-mathphys-num]{sn-jnl}

\usepackage{graphicx}%
\usepackage{multirow}%
\usepackage{amsmath,amssymb,amsfonts}%
\usepackage{amsthm}%
\usepackage{mathrsfs}%
\usepackage[title]{appendix}%
\usepackage{xcolor}%
\usepackage{textcomp}%
\usepackage{booktabs}%
\usepackage{algpseudocode}%
\usepackage{listings}%

\usepackage{float}     
\usepackage{geometry}

\theoremstyle{thmstyleone}%
\theoremstyle{thmstyletwo}%

\theoremstyle{thmstylethree}%
\usepackage{algorithm}
\usepackage{algorithmicx}
\usepackage{algpseudocode}
\usepackage{amsmath}

\floatname{algorithm}{Algorithm}

\begin{document}

\title[Article Title]{Research on the Price Prediction Algorithms 
	of Major Cryptocurrencies and a Basic Transaction Framework}

\author*[1]{\fnm{Shengjian} \sur{Chen}}\email{chshengj@mail2.sysu.edu.cn; chensj@jihualab.ac.cn}
\equalcont{La ĝenerala modelo estos eĉ pli konciza ol tio, kion mi priskribis.}

\affil{\orgdiv{Intelligent Robotics Center}, \orgname{Ji Hua Laboratory},  \city{Foshan}, \postcode{528200}, \country{China}}


\abstract{Through long-term observation and time series analysis of Bitcoin and Ethereum, we found the similarity in long-term consistent price trends, especially in the weekly K-line. We denoised and smoothed the historical data of their prices, and further derived the periodicity of their price change trends and the timing of buying and selling. We have for the first time proposed that the full application of central bank digital currencies (CBDC) is a key signal for investors to reduce their holdings of cryptocurrencies or even exit the market. In addition, we proposed momentum opening/closing prices to replace the traditional nominal opening/closing prices to accurately describe the price trends of the 24/7 financial trading market. We found that the cryptocurrency market can be regarded as a relatively independent financial market, thereby designing a safer and more efficient arbitrage strategy.}




\maketitle

\section{Introduction}\label{sec1}

Since the Bitcoin network was launched in January 2009 and the genesis block was successfully mined, the price of Bitcoin has soared from less than one ten-thousandth of a pizza to nearly $\$100,000$, with the peak once approaching $\$110,000$ \cite{BTCUSD}. Although the price of Bitcoin seems to have shown great fluctuations over the past decade or so, the continuous new highs of its market capitalization still attract many scholars and investors to study it to find out the price change patterns of Bitcoin and its internal driving mechanisms \cite{olsson2024darkflows}. The huge success of Bitcoin has driven a lot of emerging cryptocurrencies to flood into the capital market, such as Ethereum, which takes smart contracts as its core selling point, and relatively centralized platform tokens issued by major exchanges (such as Ripple) \cite{fokri2021classification}. As of March 2025, the total market capitalization of cryptocurrencies has exceeded 2.8T US dollars \cite{coinmarketcap2024}, approaching the market capitalization of Apple, the world's most valuable company. The cryptocurrency market has undoubtedly become an important force influencing the development of the world economy. Whether from the perspective of risk aversion or how to invest and make profits, the research on the price change patterns and further change mechanisms of major cryptocurrencies will provide investors with an important reference basis.

It is generally believed that the prices of cryptocurrencies fluctuate more than those of traditional financial products such as stocks, showing stronger volatility. The price of cryptocurrencies can rise or fall by $20\%$ or even more within a single day, demonstrating strong uncertainty and investment risks \cite{vasudeva2023crypto}. In order to reveal the influencing factors of the volatility of cryptocurrencies and predict their prices, many scholars have expounded the price influence mechanism from both macro and micro perspectives respectively. The mainstream view holds that the price of cryptocurrencies is at least influenced by the following factors. The most direct influencing factor is the circulation volume of cryptocurrencies. Like traditional commodities, the price of cryptocurrencies is mainly determined by the relationship between market supply and demand\cite{rudd2024forecasting}. When the supply of money in the cryptocurrency market decreases, for instance, when the mining cost increases, its price tends to rise; conversely, the price may fall. Some scholars believe that the price of cryptocurrencies is significantly affected by the government's support for them and related policies and regulations. Especially when some major countries adopt strict regulatory or even ban policies on cryptocurrencies, their prices will drop sharply. John Rile\cite{riley2021china} pointed out that the commercial policy issued by the People's Bank of China in 2013, which prohibited financial institutions from participating in Bitcoin-related businesses, once caused the price of Bitcoin to plunge by $50\%$. In addition, the overall price of the cryptocurrency market is also influenced by the global macroeconomy \cite{masiha2022effects}. When the global economy is in recession, international trade frictions intensify, and people hold a pessimistic attitude towards economic expectations, investors will increase their purchases of cryptocurrencies and use them as safe-haven assets, thereby pushing up their prices. Conversely, investors are more inclined towards traditional assets, which leads to a decline in the prices of cryptocurrencies. Internal market competition within cryptocurrencies can also lead to price fluctuations \cite{stylianou2021marketconcentration}. Cryptocurrency is a decentralized non-sovereign currency based on blockchain technology, and its most notable features are security and privacy. The maturity and innovation of encryption technology are important factors influencing its price. The improvement of technologies such as system throughput, the scalability of trading scenarios, and the programmability of trading conditions can enhance investors' confidence and expectations in cryptocurrencies, thereby driving up prices \cite{shin2022crypto}.

Some scholars believe that the prices of cryptocurrencies are influenced by a variety of complex factors, and thus adopt machine learning-based methods to predict their prices. GYEONGHO KIM  et al. \cite{kim2022deeplearning} believe that the price of cryptocurrencies is affected by market factors such as historical prices, candlestick charts, transaction size, and moving averages. They proposed a neural network model based on LSTM and the attention mechanism and adopted the CPD strategy to train a large amount of data collected from the blockchain network. The trained model can predict the price trend of Bitcoin. Gurgul \cite{gurgul2023forecasting} introduced a new method for predicting cryptocurrency prices. By analyzing news and social media content (mainly from Twitter and Reddit), it assesses the impact of public sentiment on the cryptocurrency market and uses machine learning and natural language processing technologies to predict the prices of Bitcoin and Ethereum. Feizian \cite{feizian2023sentiment} and Azamjon \cite{azamjon2023volatility} proposed a dictionary-based text analysis model, which combines the extracted text features with the weighted sentiment score features, thereby predicting the prices of Ethereum, EOS and Cardano relatively well.

Some scholars believe that although the randomness of cryptocurrency fluctuations is relatively strong, its data type is single, the data dimension is low, and the market is relatively closed. Using traditional statistical or learning methods can significantly improve the efficiency of the algorithm and provide accurate prediction results. Zhongwen Tong \cite{tong2022nonlinear} comprehensively applied techniques such as BDS test and Hurst index test, confirming that the price fluctuation of Bitcoin has nonlinear dynamic characteristics. It was found that the price fluctuations of cryptocurrencies do not follow random walks but have periodic trends and inherent long-term unpredictability. Peng Chen \cite{peng2020forecasting} developed a web application to study the fluctuation patterns of cryptocurrency prices. It mainly uses time series analysis and correlation analysis methods, allowing cryptocurrency investors to interact with the application and make investment portfolios based on the prediction results. Jain \cite{jain2018tweets} considered the differences in attributes between real currencies and cryptocurrencies and used a multiple linear regression model to predict their future prices. Abraham \cite{abraham2020patterns} applied the Johansen Test based on machine learning algorithms to study the price trends of cryptocurrency samples and check whether they are predictable. The results indicate that there is a long-term correlation between the prices of cryptocurrencies. These studies indicate that using traditional numerical analysis methods to predict the prices of cryptocurrencies has long-term stability.

Models based on machine learning often have strong generalization capabilities and can make relatively accurate result predictions for high-dimensional input data. However, its generalization ability relies on a vast amount of high-dimensional data, but it is usually difficult for us to obtain accurate values for data other than cryptocurrency prices, that is, the data type is single, and the data dimension is low. For example, the input datasets of cryptocurrency price prediction models usually include closing price and volume, parameters technical indicators, social media sentiment, and financial news articles, blockchain features, etc. Except for Closing price and volume, the other four parameters are relatively subjective and are low-frequency data, which can easily introduce incorrect values \cite{john2024survey}. Therefore, prediction models based on machine learning can often make good predictions for stable and one-sided market conditions, but they have difficulty dealing with highly volatile market conditions and multi-cycle continuous predictions, which are often the key for investors to achieve profits or losses \cite{kyriazis2020bubble}. Considering that the only public and objective high-frequency data available is the price, we adopt traditional numerical analysis methods to model the price trend of the cryptocurrency market. Moreover, the support from major countries around the world for cryptocurrencies remains weak, fundamentally failing to sustain investors' confidence in holding them for long-term investment. Most investors are in a speculative mindset of buying high and selling low. We conducted a statistical analysis of the historical price data of major cryptocurrencies and found that they do indeed have obvious speculative nature and periodicity. The price fluctuation patterns of different currencies have convergence and a master-slave relationship. Based on our research, we believe that the cryptocurrency market can be approximately regarded as an independent financial system. With accurate historical price data, we can predict price trends relatively accurately, especially the entry price, exit price and timing of the next bull market, thereby making profits. Specifically, our contributions include:

\begin{enumerate}

\item[1)] Redefine the opening price and closing price. Opening and closing prices are core parameters in the financial market, but the cryptocurrency market has no such concepts as opening and closing prices indeed: it can be traded continuously for 24 hours per day. Especially in the speculative market, the trading cycle is a very important parameter, which reflects the fluctuation cycle of investors' mentality. Therefore, we propose the concept of momentum price.
\item[2)] Obtain the prediction model for the rise and fall of major cryptocurrency prices. The long-term trend of macro prices in the cryptocurrency market is predictable. The seemingly random prices have relatively stable patterns of rise and fall. For the overall speculative financial market, we believe that its price trend satisfies the periodic mean reversion characteristic. However, due to the influence of the size of investors and inflation, its overall price is in a wavy upward trend.
\item[3)] The mutual influence law of the prices of Bitcoin and other major currencies was studied. We have found that the major cryptocurrencies share similar patterns of rise and fall. During one-sided market movements, Bitcoin almost always leads the gains, with other cryptocurrencies following suit. The fluctuation range of other cryptocurrencies is greater than that of Bitcoin.
\item[4)] Propose a robust and effective arbitrage algorithm. Point out the timing of exiting the cryptocurrency market and propose a low-frequency buy and sell combination strategy based on momentum price in combination with the rise and fall rules of cryptocurrencies. This strategy aims to achieve an annualized 3 to 5 times.

\end{enumerate}

\section{Related Works}\label{sec201}
\subsection{Speculative Nature of the Cryptocurrency Market}\label{sub2x}

The traditional view holds that the cryptocurrency market is a speculative market fraught with financial risks. Mosbey \cite{mosbey2024addiction} conducted a follow-up study on 487 cryptocurrency investors and found that this group generally suffered from varying degrees of physical and mental health harm. The sources of the harm mainly included Fear of missing out (FOMO), impulsivity, and problem gambling. Cryptocurrency investors are often in a state of high mental tension, with influencing factors including the high volatility of the cryptocurrency market, 24/7 trading, and the intensive push of returns and opportunities on social media. Under the influence of these factors, Colianni \cite{colianni2015twitter} believes that when market participants, especially retail investors, trade based on expectations of future price trends rather than the investment value or historical performance of assets, large-scale cryptocurrency speculation occurs, leading to the formation of price bubbles. Many investors do not base themselves on optimism about the value or application prospects of cryptocurrencies. Instead, under the combined influence of rumors, hype and investor sentiment, they hope to make profits through short-term price fluctuations. The sentiment in the digital currency market fluctuates greatly. Once the market shows a clear upward trend, most investors will follow the trend and enter the market. However, once some unfavorable signals appear in the market, such as the introduction of crackdown measures by some major countries or governments, a large number of cryptocurrencies will be sold off. The panic and greed of investors make the price fluctuations of cryptocurrencies extremely intense \cite{murray2025speculation}.

The virtuality of cryptocurrencies is the main reason for investors' speculative behavior. Traditional economics holds that the price of a product must be determined based on its value, especially its use value. Cryptocurrencies, however, are merely a string of electronic signals, even inferior to the currently popular credit currencies. They are not bound to any valuable items, and their prices are entirely determined by the confidence of investors \cite{bholane2025proscons}. Cryptocurrencies and central bank fiat currencies are in an incompatible competitive relationship. Currently, there are not many countries that recognize and allow the circulation of cryptocurrencies. For instance, the US Treasury Department once stated that it would adopt stricter regulatory measures for the cryptocurrency market to prevent the breeding of illegal activities such as tax evasion. The Central Bank of Turkey has suddenly shifted from accepting cryptocurrencies to ceasing their use as an effective form of payment. Mainstream cryptocurrencies are generally issued in a virtualized mining manner, and virtual currencies with no use value are endowed with "consensus value" or "co-governance value", forcing them to appear to have the pricing power of legal tender \cite{curry2025trust}. In addition, low liquidity is also a reason for the large fluctuations in cryptocurrency prices. Although research indicates that the number of global cryptocurrency holders has reached 560 million \cite{tripleadata}, $90\%$ of them are concentrated in the United Arab Emirates, Singapore, Turkey and Argentina, and their liquidity is very low. Meanwhile, the numerous secondary markets of cryptocurrencies are rife with insider trading, backroom deals and security loopholes, and are not as open, transparent and fair as they claim. The short selling mechanism is also the reason why the cryptocurrency market is full of speculative behavior. It makes speculation more unscrupulous that both the rise and fall of the market become tools for making profits, thereby increasing the speculative attribute. Liu \cite{liu2022common} and Strych \cite{strych2022margin} analyzed many speculative behaviors and risk factors in the cryptocurrency market. Through short selling and high-leverage operations, high returns can be obtained, while also bringing greater volatility to the entire market.

\subsection{Group Behavior Patterns in the Speculative Market}\label{subsec22}

The cryptocurrency market has a relatively obvious pattern of group behavior. Milka \cite{milka2020speculation} analyzed the impact of speculative behavior and thinking patterns on the cryptocurrency market. They conducted research using the sentiment index analysis method and found that there was a high correlation between the sentiment of the speculator group and the price fluctuations of cryptocurrencies. Zhang \cite{zhang2023multiscale} proposed a multi-scale regression method to explore various attributes of Bitcoin. It was found that the cryptocurrency market can be roughly divided into three trading patterns: low, medium and high frequency. Different from the medium and short-term investment attributes of the medium frequency pattern, the low frequency pattern has specific currency or long-term investment characteristics, while the high frequency pattern represents more obvious speculative behavior. Nekhili \cite{nekhili2020hedging} explored the possible speculative or hedging trading motives in the Bitcoin futures market. It tested the relationship model between the returns of Bitcoin futures and price fluctuations. Through investigating the hedging effectiveness of Bitcoin futures, it was found that traders in the Bitcoin futures market were mainly driven by speculative sentiment. Bonaparte \cite{bonaparte2022time} pointed out that the family's time horizon is the main influencing factor for investors to invest in or speculate on cryptocurrencies. It proposed a completely rational model of cryptocurrency holding tendency, proving that the longer the time range, the greater the tendency of family-based investors to own cryptocurrencies. For families, they view cryptocurrencies as a long-term investment product rather than a speculative activity. The research of the above-mentioned scholars roughly sketches out the basic behavioral patterns of investors in the cryptocurrency market, laying the foundation for us to further explore their behavioral characteristics and trading tendencies.

In the cryptocurrency market, speculative behavior is dominant while investment behavior is secondary. The differences between them can be distinguished from three aspects \cite{diaconasu2022bitcoin}. First, the motives for investment and speculation are different. Fundamentally speaking, both investors and speculators purchase cryptocurrencies in pursuit of future gains. However, the main purpose for investors to purchase cryptocurrencies lies in their confidence in the long-term rise of cryptocurrency prices and to avoid asset losses based on fiat currency due to inflation. The main purpose of speculators buying and selling cryptocurrencies is not to recognize their technology or concepts, but to focus on the short-term fluctuations in the market price of cryptocurrencies, hoping to obtain income from the price difference between buying and selling. Secondly, the trading frequencies of investment and speculation are different. Investors are willing to hold cryptocurrencies for the long term to enjoy the premium and capital gain brought by the annual price increase. They mainly focus on long-term investment returns and aim to seek long-term benefits. Therefore, investors generally pay great attention to and care about the technical level and the number of holders of the selected cryptocurrencies. Speculators are enthusiastic about the rapid turnover of transactions and seek profits from high-frequency buying and selling, and thus mainly focus on short-term gains. Speculators generally do not pay much attention to the technical level and future appreciation prospects of the selected currency, but rather focus more on trading trends such as changes in market supply and demand and shifts in hotspots. Third, the trading timing for investment and speculation is different. Investors do not immediately engage in cryptocurrency trading when the price of cryptocurrencies rises rapidly or fluctuates greatly. They only actively participate in market trading when the price drops and may cause losses or when the price is relatively stable. Speculators, on the other hand, actively participate in market trading regardless of whether the currency price rises or falls, and regardless of the methods adopted, as long as they expect to make a profit.

\section{The Basic Model of Cryptocurrency Price Fluctuations} \label{sec3x}
Given the highly speculative and volatile nature of the cryptocurrency market, we believe that the participants in this market can be roughly divided into retail investors and market manipulators. Market manipulators own a large amount of fiat currency funds and cryptocurrency assets, which can be confirmed by the fact that a large number of Bitcoins on the blockchain are held by a few accounts \cite{lehmann2019ownership}. Market manipulators can cause significant fluctuations in cryptocurrencies with a small amount of capital investment, thereby stirring up the emotions of retail investors and reaping huge profits through buying low and selling high and contract leverage. Then we need to study the behavior patterns of market manipulators and the response mode of retail investors in the cryptocurrency market, so as to obtain a general model of cryptocurrency price fluctuations and make profits by buying at low prices and selling at high prices at the right time.

\subsection{The Irreplaceability of Cryptocurrencies}\label{subsec31}

Fundamentally, the volatility of cryptocurrencies is determined by the policies and regulations of various governments. When the development of cryptocurrencies is encouraged by the government, their prices will soar, conversely, they will fall. Even in extreme cases, such as when all the world governments unanimously introduce a ban policy, their prices may directly drop to zero. However, over the past decade or so since the introduction of cryptocurrencies in 2009, they have been legally tradable in many countries, such as the United States, with their market value continuously expanding and the variety of currencies constantly increasing, remaining in a state of semi-regulation and semi-laissez-faire. One important reason why cryptocurrencies have not been truly banned, which has been overlooked by many scholars, is that they indirectly provide key technical reserves for future legal tender, central bank digital currency (CBDC).

The technical prototype verification for CBDC
The electronic features that cryptocurrencies can offer include anonymity, security and decentralization \cite{sharma2020blockchain}. Anonymity is the technical guarantee for CBDC to increase system throughput when conducting large-scale transactions. Generally, controllable anonymity is adopted for transactions, that is, small transactions are completely anonymous to improve submission efficiency, while large transactions use blockchain technology for traceability to prevent CBDC from being used for illegal activities such as telecom fraud, online gambling, money laundering and tax evasion. Security is the fundamental guarantee for CBDC to achieve full digitalization. CBDC can provide each personal account with a digital certificate based on a public-private key pair, ensuring that every transaction is a secure one signed with a private key and verified with a public key. The decentralized feature is an important technical support for the stable operation of CBDC. CBDC can adopt a distributed service deployment architecture to prevent the failure of a single central service from causing the collapse of the entire system, thereby improving the fault tolerance of the entire CBDC system \cite{sethaput2023cbdc1}.

In addition to the above three major characteristics, the industry often overlooks the key feature of cryptocurrencies, that is, programmability. Programmability was first introduced and implemented in the form of smart contracts in Ethereum, the second-largest cryptocurrency. To achieve the customizable feature of future payment scenarios, CBDC must achieve complete programmability, which is the fundamental difference between CBDC and traditional fiat currencies. Sethaput \cite{sethaput2023cbdc2} explored the application prospects of CBDC in the fields of retail, trade and supply chain finance, which allow financial transactions to be automatically executed and managed according to preset rules, significantly enhancing the efficiency of payment and settlement. Programmability, as a major upgrade and breakthrough in the financial system, lays the foundation for the fiat currency to move from offline to online and ultimately achieve the goal of full digitalization.

CBDC based on smart contracts
To achieve a CBDC that can be applied on a large scale, it is necessary to implement a mechanism that automatically executes based on payment scenarios and trigger conditions. Currently, the technical means with such characteristics are smart contracts. The terms and agreements listed in smart contracts are verifiable and automatically execute when the preset conditions are met. CBDC wallets can achieve smart payments by loading smart contracts with pre-set business logic. Some of the main technical difficulties of smart contracts have been basically overcome with the help of a lot of gained application experience and data on some cryptocurrencies (such as Ethereum). After being certified by relevant industry participants or trusted institutions, the automatic execution of CBDC smart contracts can be triggered, thereby solving the problem of frequent errors caused by traditional transactions being calculated and executed manually. For instance, in actual transaction scenarios, if a signed contract fails to be fulfilled, the untrustworthy party often abscond directly, causing property losses to the trustworthy party. The following characteristics of smart contracts effectively solve this problem.

Automatic execution: The execution of smart contracts does not rely on the participation of third-party institutions. The supervision and arbitration of the contracts are all accomplished by computer networks.
Immutability: Once a smart contract is deployed, its content cannot be tampered with, and no party can interfere with the execution of the contract alone.
Low cost: Once a breach occurs, the contract code is enforced, and its enforcement cost is lower compared to traditional contracts.
Openness and transparency: The smart contract code is made public to all network participants and can be viewed by anyone, featuring high transparency.

With the powerful functions provided by smart contracts, by loading smart contracts that do not affect the currency attributes of CBDC, conditional payment functions can be automatically triggered based on time conditions, application scenarios, locations, commodity categories, user groups, etc. This enables CBDC to have highly programmable and intelligent characteristics, allowing business rules to drive the flow of funds and making the fund flow controllable and traceable, thus, becoming the true next-generation smart currency. This is the financial means that must be possessed to realize a smart city. Figure 1 shows the general process of online transactions using smart contracts.

\begin{figure}[H]
	\centering
	\includegraphics[width=1.0\textwidth]{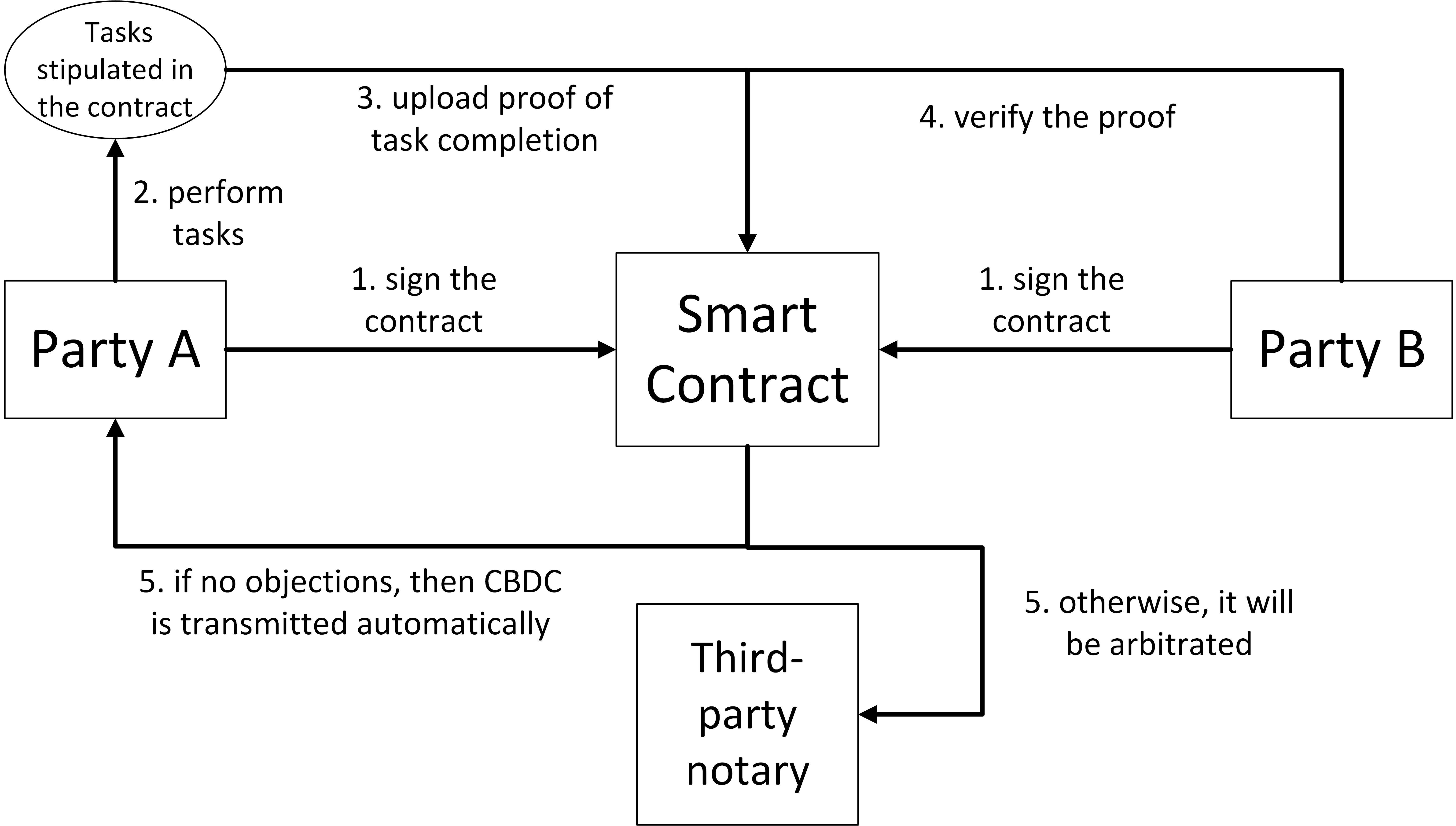}
	\caption{The general process of smart payment based on CBDC}
	\label{fig:fig1}
\end{figure}

\subsection{Selection of Investment Currency} \label{subsec32}

Given the highly speculative nature of the cryptocurrency market, we only select cryptocurrencies with large market capitalization, a large number of holders, core technological support, and relatively small volatility for investment, to ensure stable returns rather than high-risk ones. We compared ten mainstream cryptocurrencies, and in the end, only Bitcoin and Ethereum, the two cryptocurrencies with the largest market capitalization, met the requirements. The limited issuance of Bitcoin and its high mining difficulty make it be the belief in the hearts of cryptocurrency investors, almost determining the overall rise and fall trend of the cryptocurrency market. Ethereum also has the characteristics of limited total supply and high mining difficulty. And it has creatively proposed smart contract technology and relies on virtual machines to run automatically. It is the key technology for large-scale promotion of future fiat CBDC, and this technology is sufficient to make Ethereum comparable to Bitcoin. Apart from these two currencies, all other cryptocurrencies are merely simple technical replicas, highly speculative in nature, and lack long-term investment value. Figure 2 shows the maximum price multiple of the top 10 cryptocurrencies by market capitalization, defined as the ratio of each day?s closing price to the lowest price observed since August 2017. It can be seen that LINK?s price multiple reaches up to 500,000x. Even excluding the most volatile LINK and BNB, as can be seen from Figure 3, the maximum price multiple of other mainstream cryptocurrencies can still be as high as over 1,000x, which is highly speculative and lacks long-term investment value. In addition, due to the high degree of opacity of this market, we do not engage in contract trading. We only do spot trading.

\begin{figure}[H]
	\centering
	\includegraphics[width=1.0\textwidth]{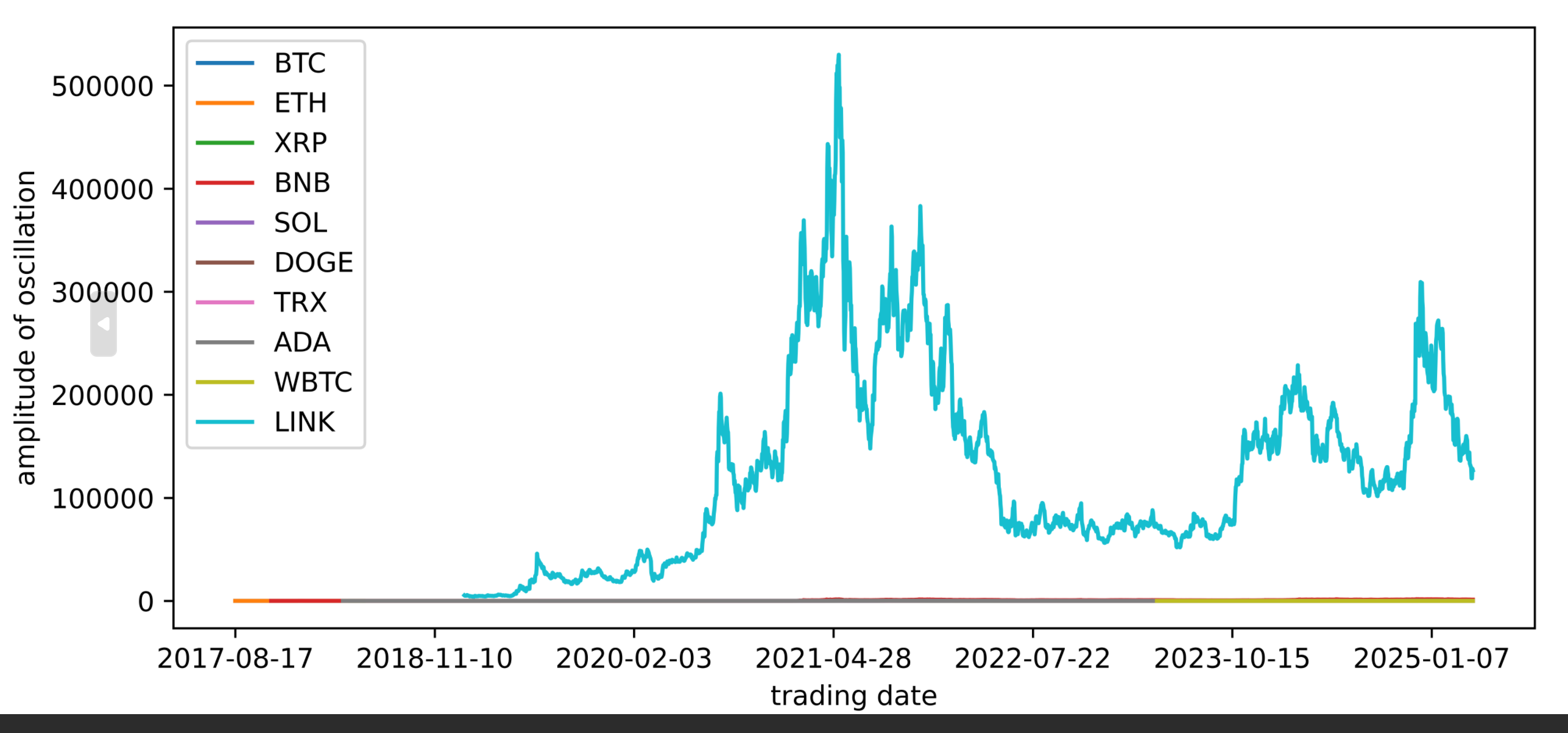}
	\caption{ The fluctuation range of major cryptocurrency prices}
	\label{fig:fig2}
\end{figure}

\begin{figure}[H]
	\centering
	\includegraphics[width=1.0\textwidth]{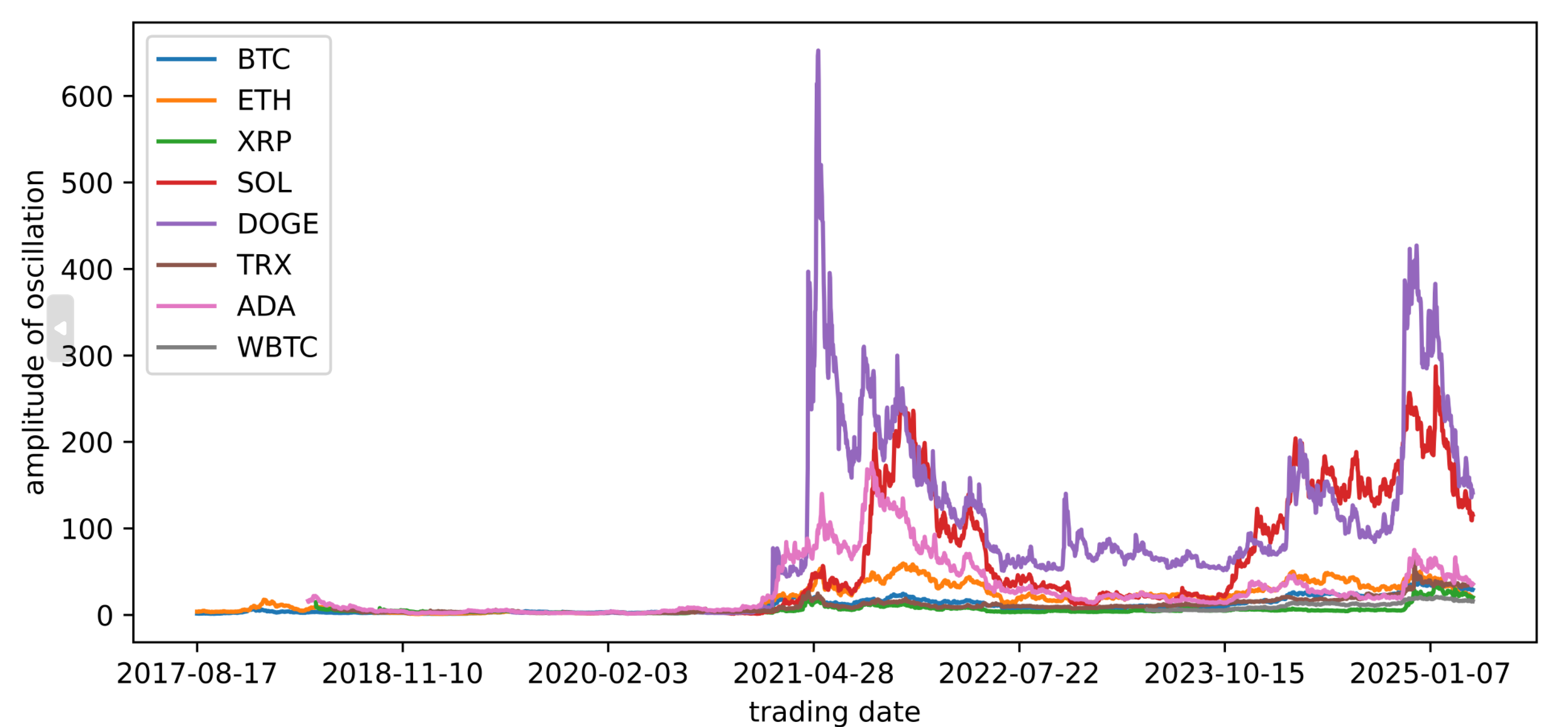}
	\caption{ The fluctuation range of prices of major cryptocurrencies excluding Link and BNB}
	\label{fig:fig3}
\end{figure}

\subsection{Timing of Exiting the Cryptocurrency Market} \label{subsec33}

The most important question in trading financial assets in the speculative market is when one should exit the market. As can be seen from the analysis in Section 3.1, cryptocurrencies can provide the necessary technical support for CBDC, and the relevant technologies can be slightly modified and transplanted to the research and development of CBDC. At present, apart from smart contract, there are no technical or application challenges that need to be overcome for cryptocurrencies. Due to the characteristic that smart contract cannot be tampered with once deployed, once errors are found in the contract content, how to withdraw it or avoid losses as much as possible remains a difficult problem to solve. Therefore, research related to smart contract is still ongoing, and the application in cryptocurrencies is conducive to the improvement of its theory and practice, thereby providing key R\&D experience for CBDC. Based on this, we believe that as long as the CBDC-related technologies are not fully ready, the cryptocurrency market will not be completely banned, and the prices of mainstream cryptocurrencies will still remain at a relatively high level. However, since cryptocurrencies and fiat currencies are in an incompatible competitive relationship, if CBDC have fully matured, especially when they are about to be widely promoted, they will exert pressure on cryptocurrencies. As a result, the possibility of cryptocurrencies being banned will increase significantly. This is the time for investors to consider withdrawing from the cryptocurrency market. The specific algorithm for order cancellation varies depending on the timing, position size and the amount of loss that can be tolerated.

We used the currently popular artificial intelligence model, LLM \cite{yao2024llm} to train the sample data collected on the Internet regarding cryptocurrencies, CBDC, government policies, smart contracts, etc. The goal is to obtain the credibility $cd \in [0, 1]$  of the upcoming full-scale promotion of CBDC. The relevant model training methods can be referred to in these references \cite{lei2025bitcoinllm,chaffard2025forecasting}. We deploy the trained model to a high-performance server, and then send the real-time sample data collected from the Internet into the AI model, and its prediction results are fed back to the function module $CancellAll$. 
\begin{itemize}
\item When $cd$ exceeds a certain value but is less than the set risk control warning factor $LTH$, for instance, when the People's Bank of China starts to issue CBDC, we need to be vigilant, but there is no need to adjust the position for the time being. 
\item When $cd$ exceeds LTH but is lower than the high threshold $HTH$, such as when the European Central Bank develops and issues CBDC, it indicates that the risk of cryptocurrencies being banned is increasing, and some buy orders for Bitcoin and Ethereum need to be withdrawn. In Algorithm 1, we sort all the buy orders  and  for Bitcoin and Ethereum in ascending order by price. Then, we remove the orders with a proportion of  $\displaystyle (\frac{LTH}{cd}+\frac{cd}{HTH})/2$ on the front of these two lists, and re-place the exchanged USDT at the expected prices  and  for Bitcoin and Ethereum respectively. 
\item If the credibility $cd$ provided by the AI model is higher than $HTH$, for example, when the Federal Reserve starts issuing CBDC, then we think the risk of the cryptocurrency market being banned will increase significantly. We need to cancel all buy and sell orders for Bitcoin and Ethereum, and sell all the Bitcoin and Ethereum we own at the current price of  $C_{btc}$ and $C_{eth}$.
\end{itemize}
 
Since we adopt a prudent investment strategy, aiming for high returns while preserving our capital as much as possible, this step of order cancellation is of great significance. When necessary, manual order cancellation can be carried out to prevent losses caused by the inconsistency between the model's predicted results and the actual situation.

  \begin{algorithm}
	\caption{Urgent handling of pending orders}
	\begin{algorithmic}[1] 
		\Require $LTH, HTH, E_{btc}, C_{btc}, Bo_{btc}, So_{btc}, E_{eth}, C_{eth}, Bo_{eth}, So_{eth}$
		\Ensure bool
		\Function {CancelAll}{$ $}
		\State $cd \gets$ \Call{AIPredition}{$ $}
		\If {$ cd <= LTH$}
		\State \Call{Ignore}{$ $}
		\State return \textbf{False}
		\ElsIf {$LTH < cd < HTH$}
		\State $usdt \gets$ \Call{Delete}{$Bo_{btc}, (LTH/cd+cd/HTH)/2, Asc$}
		\State \Call{Place}{$usdt, E_{btc}, Bo_{btc}$}
		\State $usdt \gets$ \Call{Delete}{$Bo_{eth}, (LTH/cd+cd/HTH)/2, Asc$}
		\State \Call{Place}{$usdt, E_{eth}, Bo_{eth}$}
		\State return \textbf{False}
		\Else
		\State $btc,eth \gets$ \Call{DeleteAll}{$Bo_{btc},So_{btc},Bo_{eth},So_{eth}$}
		\State \Call{Place}{$btc, C_{btc}, So_{btc}$}
		\State \Call{Place}{$eth, C_{eth}, So_{eth}$}
		\State return \textbf{True}
		\EndIf
		\EndFunction
	\end{algorithmic}
\end{algorithm}

\subsection{General Patterns of Price Fluctuations} \label{sub34}

In a speculative market, price fluctuations are always cyclical, and the cryptocurrency market is no exception. A typical speculative cycle usually begins with a relatively low price point, relying on the continuous psychological game between buyers and sellers to advance, thereby generating a large number of price fluctuations, big or small, until the price soars to a certain peak and causes a large amount of selling and ends. During an investment cycle in the cryptocurrency market, the price can increase by several times or even tens of times. As long as people's nature of pursuing benefits and fearing losses remains unchanged, this cycle will continue indefinitely. The major cycle of cryptocurrencies is generally driven by market manipulators absorbing a large number of cryptocurrencies at low prices, and then continuously placing orders to sell at the price peak. As the purchase volume of retail investors changes, the price is constantly adjusted. That is, when retail investors buy in large quantities, the price of cryptocurrencies is raised; when they sell in large quantities, the price is lowered. Consequently, most retail investors fail to hold onto cryptocurrencies long enough to capture the peak price. They are always buying and selling frequently under the influence of panic, thus forming a lot of fluctuations. That is to say, the price curve of cryptocurrencies as a whole shows a distinct cyclical oscillation characteristic, but within a major cycle, there are many peaks and troughs, and between each peak and trough, there are many smaller fluctuations.

\begin{figure}[H]
	\centering
	\includegraphics[width=1.0\textwidth]{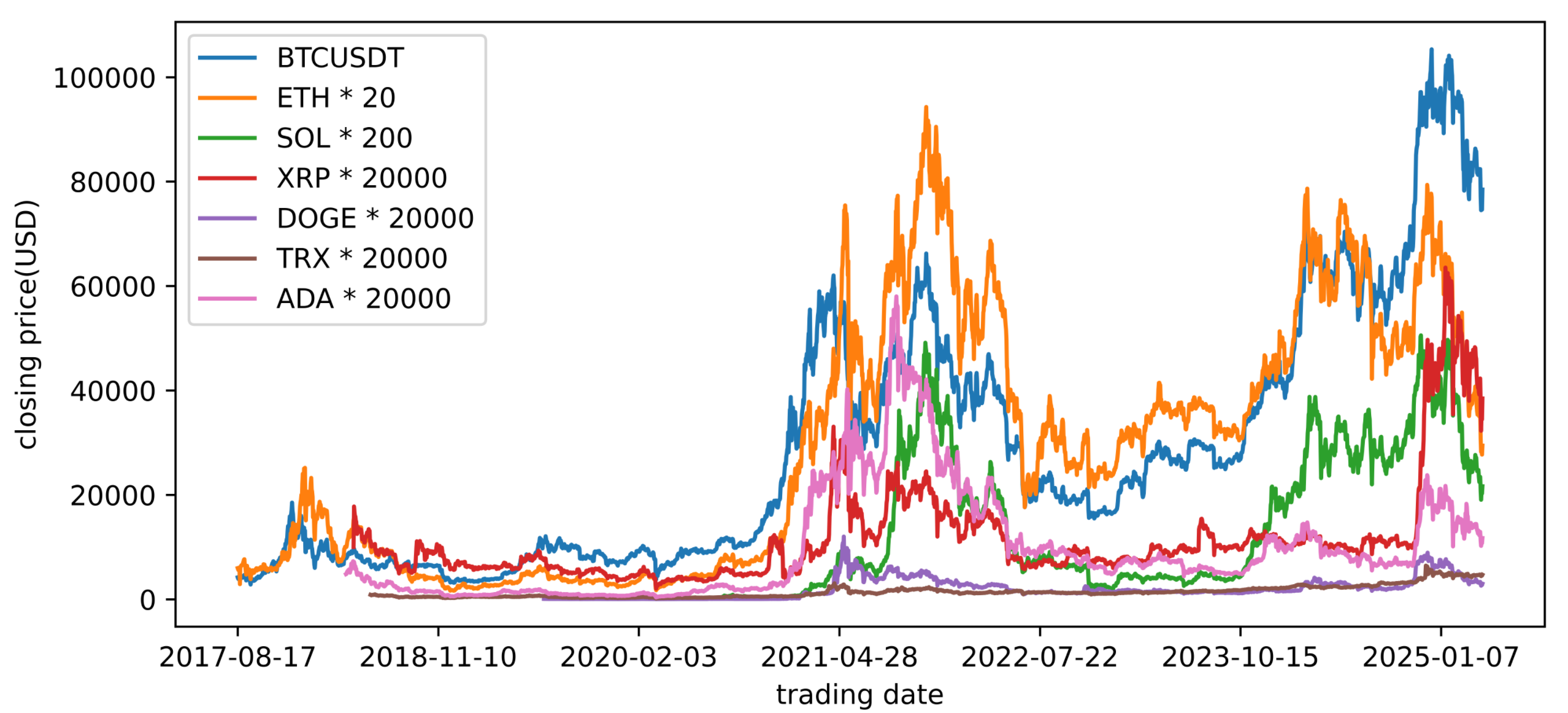}
	\caption{ The convergence of daily prices of major cryptocurrencies}
	\label{fig:fig4}
\end{figure}

Figure~\ref{fig:fig4} illustrates the daily opening prices of the ten largest cryptocurrencies by market capitalization over the past eight years. All of these assets exhibit frequent price swings; the smaller-cap coins tend to oscillate more often but with lower amplitude, and their movements appear essentially random. However, as long as we eliminate the interference of these small fluctuations, we can see that almost all the major cryptocurrencies on the chart have a consistent trend, especially when we multiply the prices of currencies other than Bitcoin by different multiples (for example, ETH expands by 20 times and SOL expands by 200 times). We can observe that all currencies have formed two consistent major cycles between 2017-2018 and 2021-2022. That is, in the speculative market of cryptocurrencies, due to the influence of various macro factors, such as economic development and inflation, the relationship between the two cycles is not a simple repetitive one, but a stepwise growth one. This provides a strong practical basis for us to hold cryptocurrencies for the long term. 

\begin{figure}[H]
	\centering
	\includegraphics[width=1.0\textwidth]{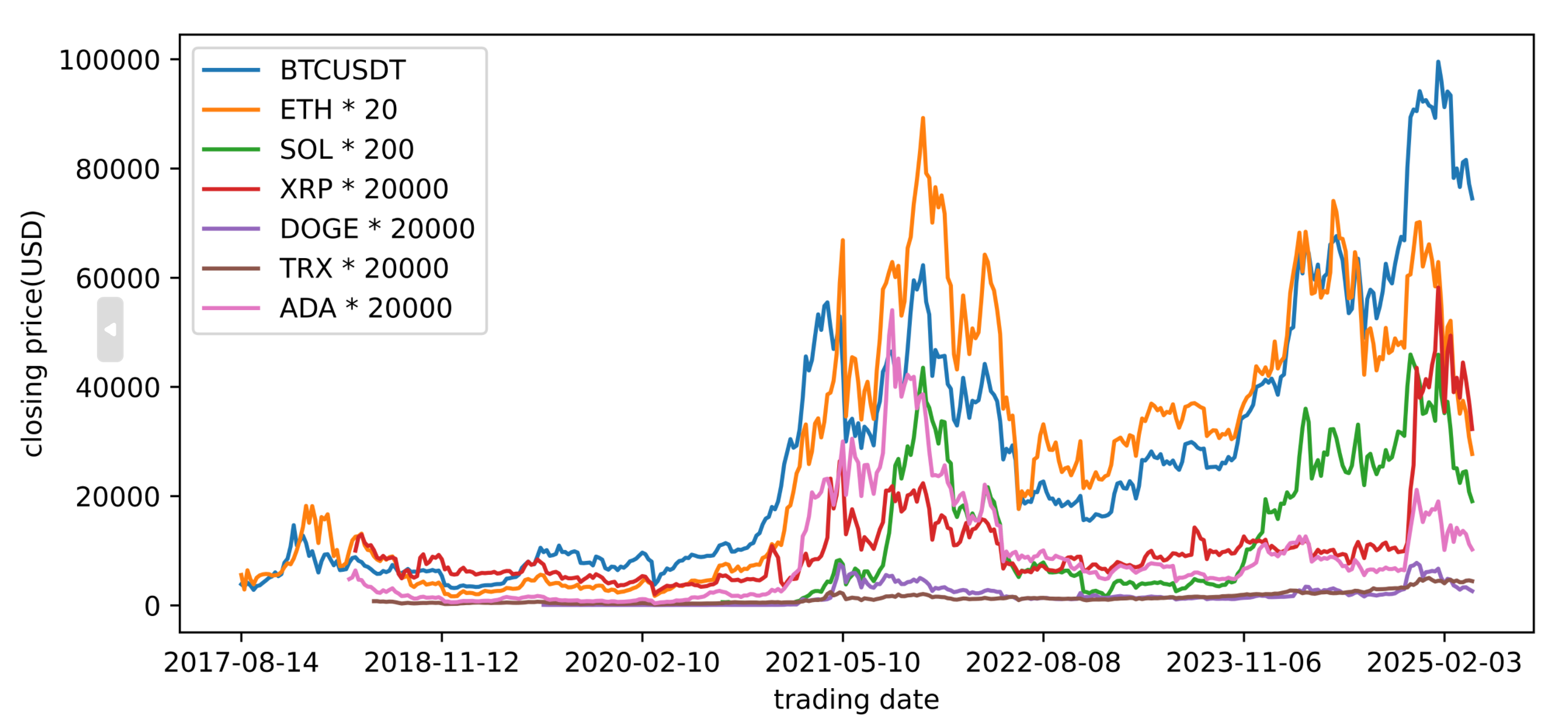}
	\caption{The convergence of weekly prices of major cryptocurrencies}
	\label{fig:fig5}
\end{figure}

Overall, we need to accurately identify the intentions of the market manipulators, find some stable price trends, not be affected by short-term price fluctuations in our emotions, and make wise investment decisions. Even in a stable price trend of the cryptocurrency market, the price difference between small peaks and troughs can be as high as over $50\%$, but we don't need to pay attention to these small fluctuations. A $5\%$ fluctuation and a $50\%$ fluctuation are essentially the same. Once a trend is formed, prices can soar several to dozens of times. Once it rises close to its peak, market manipulators begin to sell off in large quantities when they find that there are enough retail investors to fully absorb the cryptocurrencies in their hands. At this point, liquidity has increased significantly. Once the upward trend starts to reverse, the rate of decline will be faster than the initial increase. The cryptocurrency market keeps repeating the same ups and downs cycle under the continuous low-price absorption and high-price selling by major market manipulators, but each new cycle always starts anew at a higher price. The weekly opening prices shown in Figure~\ref{fig:fig5} better depict the general pattern of the rise and fall cycle of cryptocurrencies. Other major cryptocurrencies fluctuate roughly along with the rise and fall of Bitcoin, with a slight delay in time, but the extent of the rise and fall is greater. The main reason why we choose the weekly K-line chart is that it takes time for the market manipulator to stir up the emotions of the retail investors. The weekly K-line chart can better reveal the collective behavior patterns of the retail investors under the manipulation of the major market manipulator.

\section{Optimization of Model Parameters} \label{sec4}

From the price rules of cryptocurrencies derived in Chapter 3, we have summarized the general investment strategy, which has three key points. The first is to only select cryptocurrencies with a large number of holders and core technological support, namely Bitcoin and Ethereum, with Bitcoin as the main currency and Ethereum as the secondary one. Second, once it is confirmed that all the core technologies required for CBDC have been mastered and are about to be widely promoted, one should start to significantly reduce holdings or even completely withdraw from the cryptocurrency market. Thirdly, it is necessary to identify the low and high points of each major cycle of cryptocurrencies, so as to make profits by buying low and selling high. In this chapter, we will conduct a quantitative study on the price model of cryptocurrencies to obtain the rise and fall patterns and buying and selling timing between Bitcoin and Ethereum. For instance, through the analysis of historical data and the summary of market experience, it is generally found that each major cycle is initiated by Bitcoin first, which then drives up the prices of other cryptocurrencies. How long after buying Bitcoin is it appropriate to purchase Ethereum? How long does it take for each major peak to reach a major trough? To obtain a high-precision price model, we need to solve the following two problems. The first is to determine which parameter (such as the opening price, closing price, etc.) should be used as the main basis for the cryptocurrency K-line chart. The second is to determine the approximate price range and occurrence time of each major peak and trough on the K-line chart.

\subsection{Nominal Price and Momentum Price} \label{sub41}

Obtaining accurate price trends is the key to making profits in the cryptocurrency market. The opening and closing prices on the K-line chart are the first-hand information for understanding the financial market and are the key parameters for investors to understand the price trends in each major cycle. They reflect the behavior and sentiment fluctuations of most investors during this cycle. In traditional markets such as stocks, each trading day has fixed opening and closing times. After the market closes, investors have sufficient time to review the day's market conditions and trends, and their investment impulses can be calmed down. Therefore, it is appropriate to determine a small investment cycle based on the traditional opening and closing times. However, in the cryptocurrency market, there are no concepts of opening and closing. It can be traded 24/7 and cannot simply copy the traditional definitions. Investors' behaviors and emotions in buying and selling cryptocurrencies vary in duration and fluctuation range each day. It is inappropriate to set 0:00 and 24:00 as the opening and closing times of each day, and the corresponding opening and closing prices cannot fully reflect the behavioral and emotional fluctuation trends of investors within a small trading cycle. Establishing a model based on the traditional opening and closing prices to predict the price trend of cryptocurrencies will lead to significant errors. Therefore, it is necessary to find a more suitable parameter to optimize the model.

Momentum strategy is a common strategy in the stock trading market and a methodology for capturing trading trends. It is often referred to as the "inertia effect" and was proposed by Jegadeesh and Titman in 1993~\cite{jegadeesh2011momentum}. Their research found that the return rate of stocks has a tendency to continue the original movement direction, that is, stocks with a relatively high return rate in the past period will still achieve a return rate higher than the average in the future. We draw inspiration from the research on momentum strategies. Whether it is a major cycle (the duration between two bull markets in cryptocurrencies) or a small cycle (such as a trading day), the market state is nothing more than volatility and trends. Then, once the trading activities at the closing time of a trading day and the opening time of the previous trading day have momentum characteristics, we can then consider that the current closing price cannot reflect the trend characteristics of the current trading day. Therefore, we propose the momentum opening and closing price to reflect the trading trend within a small trading cycle.

\paragraph{Nominal opening and closing price} 
The period from 00:00:00 to 23:59:59 local time is a natural trading day. 00:00:00 is the nominal opening time, denoted as $t^{no}$, and its determined opening price is the nominal opening price, denoted as $p^{no}$. 23:59:59 is the nominal closing time, denoted as $t^{nc}$, and its determined closing price is the nominal closing price, denoted as $p^{nc}$.

\paragraph{Momentum analysis} 
We provide the definitions of the momentum opening price $p^{mo}$ and the momentum closing price $p^{mc}$ here. The arrow symbol is used to indicate the direction of time movement. The left arrow indicates a movement earlier than the nominal opening time of the day, and the right arrow indicates a movement later. Let $t^{\rightarrow}$ be a certain time after the opening of the day, satisfying $t^{no} < t^{\rightarrow} \ll t^{nc}$. It divides all the trading prices of the day into two sequences according to time. The sequence with the earlier time is denoted as $p^{\rightarrow}_{no}$  and the quantity of elements is $|p^{\rightarrow}_{no}|$, and the sequence with the later time is denoted as $p^{\rightarrow}_{nc}$ and the quantity is $|p^{\rightarrow}_{nc}|$. Similarly, we take another fixed time $t^{\leftarrow}$ that satisfies $t^{no\leftarrow} \ll t^{\leftarrow} < t^{nc\leftarrow}$, which divides all the transaction prices of the previous day into two sequences according to time. The sequence with the earlier time is denoted as $p^{\leftarrow}_{no}$  and the quantity is $|p^{\leftarrow}_{no}|$, and the sequence with the later time is denoted as $p^{\leftarrow}_{nc}$ and the quantity is $|p^{\leftarrow}_{nc}|$, where $t^{\leftarrow}_{no}$ and $t^{\leftarrow}_{nc}$ are respectively the nominal opening and closing time of the previous day. The $i$-th element of the sequence $p^{\rightarrow}_{no}$  is denoted as $p^{\rightarrow}_{no}[i]$  , and the same applies to other sequences.

\paragraph{Condition 1 (Trend ahead):}

\[
\left\{
\begin{aligned}
	|M^{\leftarrow} - M^{\rightarrow}| &< M_\Delta \\
	\min(|M^{\leftarrow}|, |M^{\rightarrow}|) &> M_\alpha
\end{aligned}
\right.
\]

If the above conditions are met, we consider that the small cycle of cryptocurrency trading has actually begun since time $t^{\leftarrow}$ of the previous day. Then, the momentum opening price can be regarded as the transaction price at time $t^{\leftarrow}$, denoted as $P^{(t\leftarrow)}$. To eliminate the errors caused by some false transactions, the average of several transaction prices after time  can be taken to replace $P^{(t\leftarrow)}$ in practical application.

$M^{\leftarrow}$ and $M^{\rightarrow}$ are the weighted momenta proposed to describe the trend direction and the degree of rise and fall over a certain period. Let the weighted momentum within period  \( (t^{\leftarrow}, t^{no}) \) before the opening be \[
M^{\leftarrow} = \frac{ \sum_{i=1}^{|p_{nc}^{\leftarrow}| - 1} \langle p_{nc}^{\leftarrow}[i], p_{nc}^{\leftarrow}[i+1] \rangle }{ |p_{nc}^{\leftarrow}| - 1 }
,\] and the weighted momentum within period \( (t^{no}, t^{\rightarrow}) \) after the opening be \[
M^{\rightarrow} = \frac{ \sum_{i=1}^{|p_{no}^{\rightarrow}| - 1} \langle p_{no}^{\rightarrow}[i], p_{no}^{\rightarrow}[i+1] \rangle }{ |p_{no}^{\rightarrow}| - 1 }.
\]

 Additionally,  $M_{\alpha}$ and $M_{\Delta}$  are positive constants and can be determined based on trading volume and trading price in practical applications. The operation  \( \langle p[i], p[i+1] \rangle \) is called the trend strengthening operation, any function that can magnify its difference can be selected. Generally, we define it as a quadratic function:
 \[
 \langle p[i], p[i+1] \rangle =
 \begin{cases}
 	(p[i] - p[i+1])^2, & \text{if } p[i] < p[i+1] \\
 	-(p[i] - p[i+1])^2, & \text{if } p[i] \geq p[i+1]
 \end{cases}
 \]

That is, when the price rises, the trend strengthening operation is positive, and the larger the price difference, the more obvious the trend strengthening. On the contrary, when the price drops, the trend strengthening operation is negative. When the sum of the weighted momenta over a period of time is positive, it indicates that the overall trend of this period is rising; otherwise, it is falling. If the difference between the weighted momenta $M^{\leftarrow}$ and $M^{\rightarrow}$  is small and the absolute values of both are large enough, we consider that these two periods of time are actually in the same momentum period. Therefore, the logical opening price, that is, the momentum opening price, is set as the starting price of this period, $P^{(t\leftarrow)}$.

\paragraph{Condition 2 (Trend delay):}

\[
\left\{
\begin{aligned}
	\max(|M^{\leftarrow}|, |M^{\rightarrow}|) &< M_\beta \\
	|M^{\rightarrow \rightarrow}| &> M_\gamma
\end{aligned}
\right.
\]

If the above conditions are met, we consider that the small cycle of cryptocurrency trading begins at time $t^{\rightarrow}$ on that day. Then, the momentum opening price can be regarded as the transaction price at time $t^{\rightarrow}$, denoted as $P^{(t\rightarrow)}$. Similarly, in order to eliminate the errors caused by some false transactions, the average of several transaction prices after time $t^{\rightarrow}$ can be taken.

Let the weighted momentum within the time period \( (t^{\rightarrow}, t^{nc}) \)   after the opening be \[
M^{\rightarrow \rightarrow} = \frac{ \sum_{i=1}^{|p_{nc}^{\rightarrow}| - 1} \langle p_{nc}^{\rightarrow}[i], p_{nc}^{\rightarrow}[i+1] \rangle }{ |p_{nc}^{\rightarrow}| - 1 }.
\] If its value exceeds the positive constant $M_\gamma$, we think that a continuous trend has formed within the time period \( (t^{\rightarrow}, t^{nc}) \). If there is no obvious trend in the previous time period $[t^{\leftarrow}, t^{\rightarrow}]$, that is, the absolute values of $M^{\leftarrow}$ and $M^{\rightarrow}$  are both less than the constant $M_\beta$, then we think that the momentum opening price of the current day has been delayed. The constants $M_\gamma$ and $M_\beta$  need to be adjusted according to the trading volume. Generally, $M_\beta \ll M_\gamma < M_\alpha$.

\paragraph{Momentum opening and closing price}
Based on the above analysis, compared with the nominal opening price, the momentum opening price may be advanced or delayed. Correspondingly, if the momentum opening price of the current day and the next day are determined, then the momentum closing price of the current day is also determined. Its price can be taken as the price before the momentum opening time of the next day. Under normal circumstances, we only need to take one of the opening or the closing price curve to better depict the market price trend of cryptocurrencies. We mainly focus on the momentum opening price. Then the mathematical expression of the momentum opening price is:
\[
p^{mo} = 
\begin{cases}
	P^{(t^{\leftarrow})}, & \text{if condition 1 is satisfied} \\
	p^{no}, & \text{if both condition 1 and 2 are not satisfied} \\
	P^{(t^{\rightarrow})}, & \text{if condition 2 is satisfied but not 1}
\end{cases}
\]

\begin{figure}[H]
	\centering
	\includegraphics[width=1.0\textwidth]{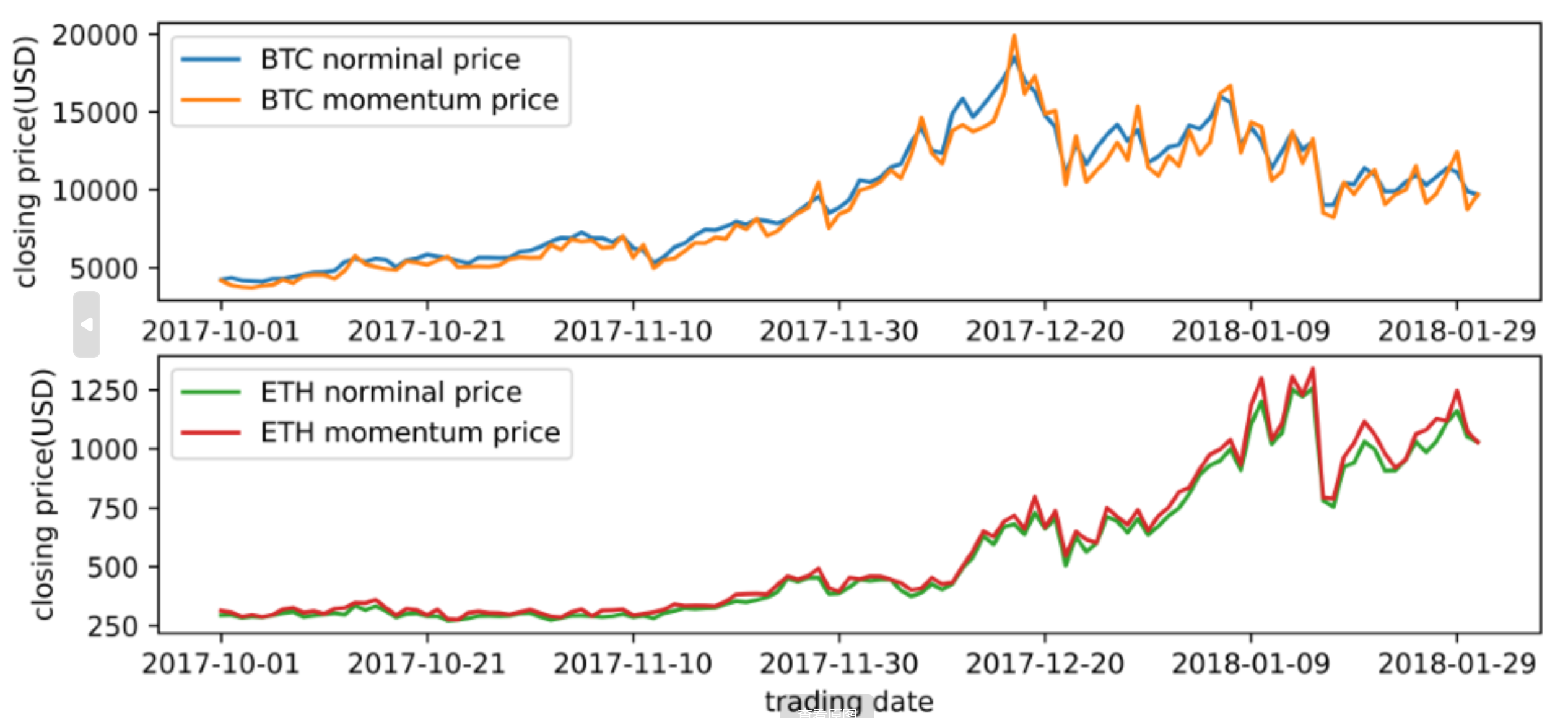}
	\caption{Comparison between the nominal opening price and the momentum opening price}
	\label{fig:fig6}
\end{figure}

Figure~\ref{fig:fig6} respectively presents the time curves of the nominal opening price and the momentum opening price of Bitcoin and Ethereum. It can be seen that the momentum curve can make the one-sided market trend more obvious, and at the same time make the sideways movement more stable, playing a role in strengthening the trend, thereby providing assistance for us to make more accurate predictions about the price trend.

\subsection{Model Feature Points} \label{sub42}

Our data source is third-party institutions such as exchanges whose trading data cannot fully reflect the overall situation of the cryptocurrency market, and there may even be a certain degree of false trading. However, its trading prices, especially the momentum prices we have adjusted, can reflect the overall trend of cryptocurrency prices. Therefore, we mainly predict the price trend based on price data and supplemented by data such as transaction volume. Determining trading feature points based on the momentum price curve is the key to making profits in the cryptocurrency market, as it reveals the main buying and selling directions of market manipulators and retail investors.

\paragraph{Troughs and peaks of the major cycle}
In the speculative market, it takes some time for a market manipulator to buy a large number of cryptocurrencies at a low price and for most retail investors to take over and eventually lead to a price crash. This period usually lasts for 2 to 5 years. Based on historical experience, the minimum and maximum values of cryptocurrencies during this period can differ by at least ten times. For currencies other than Bitcoin and Ethereum, the price difference can even reach hundreds of times or more. This is determined by the highly speculative nature of cryptocurrencies. The stock price depends on the operating performance, credit rating, dividend distribution status, development prospects and expected return level of the issuing company. Unlike traditional financial products such as stocks, the price of cryptocurrencies is entirely determined by investors' beliefs in their prospects and technology. Under the manipulation of market manipulators, a large number of retail investors are extremely prone to waver in confidence due to market volatility and cannot hold cryptocurrencies for a long time. As a result, market manipulators can easily obtain sufficient money to manipulate market prices more easily, and this cycle repeats itself.

Let the trough price of the previous major cycle be \( P^{(L\leftarrow)} \), the peak price be \( P^{(H\leftarrow)} \), and the peak trading volume be $S^{(\leftarrow)}$. Then let the trough price before the previous major cycle be \( P^{(L\leftarrow\leftarrow)} \), the peak price be  \( P^{(H\leftarrow\leftarrow)} \), and the peak trading volume be $S^{(\leftarrow\leftarrow)}$. The arrow symbol indicates the direction of time movement relative to the current cycle.

Define the peak-to-trough ratios:

\begin{align*}
	\lambda^{(1)} &= \min\left( \frac{P^{(H\leftarrow)}}{P^{(L\leftarrow)}}, \frac{P^{(H\leftarrow\leftarrow)}}{P^{(L\leftarrow\leftarrow)}} \right), \\
	\lambda^{(2)} &= \max\left( \frac{P^{(H\leftarrow)}}{P^{(L\leftarrow)}}, \frac{P^{(H\leftarrow\leftarrow)}}{P^{(L\leftarrow\leftarrow)}} \right)
	\bigg/ \ln\left( \frac{\max(S^{(\leftarrow)}, S^{(\leftarrow\leftarrow)})}{\min(S^{(\leftarrow)}, S^{(\leftarrow\leftarrow)})} \right) \\
	\lambda^{(\min)} &= \min(\lambda^{(1)}, \lambda^{(2)}) \\
	\lambda^{(\max)} &= \max(\lambda^{(1)}, \lambda^{(2)})	
\end{align*}

  where function $\ln$ is the natural logarithm. Then we predict $\lambda$, the peak-to-trough ratio of the next major cycle is most likely to satisfy:
\[
\lambda \in \left( \lambda^{(\min)}, \lambda^{(\max)} \right)
\]

The reason why we only select the previous two major cycles here is that the cryptocurrency market was highly speculative in the early stage, and the market has only tended to be relatively stable in the past few years. Moreover, considering human greed, the lowest point of the new major cycle is much lower than the highest point $P^{(H\leftarrow)}$ of the previous major cycle, but higher than $P^{(H\leftarrow\leftarrow)}$. That is, the lowest point price $P^{(L)}$ of the upcoming new cycle is:
  \[
  P^{(L)} \in \left( P^{(L)(\min)}, P^{(L)(\max)} \right)
  \]

where $P^{(L)(\max)} = \max\left( \frac{P^{(H\leftarrow)}}{\varepsilon}, P^{(H\leftarrow\leftarrow)} * \zeta \right)$ and $P^{(L)(\min)} = \min\left( \frac{P^{(H\leftarrow)}}{\varepsilon},  P^{(H\leftarrow\leftarrow)} * \zeta \right)$,  with $\zeta$ and $\varepsilon$  being trough oscillation factors, satisfying  $0 < \zeta < \varepsilon$. The typical values are $\zeta = 2$ and $\varepsilon = 4$ for Bitcoin. Similarly, we predict that the price peak  of the new major cycle will be:
\[
P^{(H)} = \lambda * P^{(L)}
\]

\paragraph{Signs of major cycle replacement}
After a major cycle comes to an end, the price of cryptocurrencies will experience a continuous and rapid decline. Within as short as two weeks and as long as several months, the price can drop several times. This is manifested on the price curve as having many segments with relatively large slopes during this period, and within these small segments, there are a few segments of price pullbacks. During this process, market manipulators sell off their cryptocurrencies in large quantities in batches, and retail investors no longer have enough buying power for the cryptocurrencies, thus leading to a panic selling wave. Let the total duration of this process be \( T^{(L)} \), and it consists of $N$ segments in total (in practice, each segment is usually one week). The price sequence formed by the momentum opening prices of each segment is $\{p^{(\downarrow)}[0],\ p^{(\downarrow)}[1],\ p^{(\downarrow)}[2],\ \ldots,\ p^{(\downarrow)}[N]\}$. Let the slope between points  $p^{(\downarrow)}[0]$ and $p^{(\downarrow)}[1]$ be $slope(p^{(\downarrow)}[0], p^{(\downarrow)}[1])$, if the following conditions are met:

\[
\left\{
\begin{aligned}
	\text{slope}(p^{(\downarrow)}[0],\ p^{(\downarrow)}[N]) &< \theta \\
	\text{count}\left( i \left| 
	\frac{\text{slope}(p^{(\downarrow)}[i],\ p^{(\downarrow)}[i+1])}{\text{slope}(p^{(\downarrow)}[0],\ p^{(\downarrow)}[N])} > \vartheta
	\right. \right) &\geq \xi \cdot N
\end{aligned}
\right.
\]

Then we believe that the market has begun to fall sharply, moving from the peak of an old cycle to the trough of a new cycle. This is manifested on the price curve as having more than half of the sharp decline segments, and the slopes of these segments are greater than that of the entire curve where $\theta < 0$, $\vartheta > 1$  and $\xi > 0.5$. The function $count$ represents counting the results that meet its parameter conditions. If price $p^{(\downarrow)}[N]$ is close to the predicted low point of this new cycle at this time, it is advisable to buy large quantity of cryptocurrencies; otherwise, one can continue to wait for the price to fall. If investors have sufficient emotional control ability, they can purchase a small number of cryptocurrencies at this time, because the price will briefly rise before hitting the bottom, thus making short-term profits.

\subsection{Overall Algorithm Framework} \label{sub43}

From the above discussion, we have derived four fundamental conditions for making profits in the speculative market of cryptocurrencies. The first is to determine the conditions for a stepwise withdrawal from the market, that is, once CBDC is officially issued in major countries around the world, investors must be aware of the urgency of exchanging the cryptocurrency back for fiat currency. Second, the momentum price curve is used instead of the nominal price curve to make the price trend more in line with the characteristics of the 24/7 trading cryptocurrency market. The third is to estimate the price troughs and peaks of the new major cycle to determine the timing of buying and selling. Fourth, determine the characteristics of the price curve when the old and new cycles transition, that is, the occurrence of a one-sided sharp decline, so as to prepare for bottom-fishing. Then our algorithm process becomes relatively clear. We buy at the bottom, sell at the peak, withdraw before the danger signal appears, and use traditional quantitative trading strategies to handle the small fluctuations, such as momentum trading strategy~\cite{chan2021quantitative,xie2023quantamental}, mean reversion strategy~\cite{wood2021slowmomentum1,wood2021slowmomentum2}, etc. Then we can basically ensure that we can make a profit of 2 to 10 times within a major cycle without losing the principal.

Algorithm 2 provides robust strategies for buying and selling physical Bitcoin and Ethereum.

\begin{itemize}
	\item First, we convert the nominal price $p_{no}$ to the momentum price $p_{mo}$  in real time and add it to the price sequence $P_{btc}$ and  $P_{eth}$ of Bitcoin and Ethereum.
\item Second, based on the momentum price weekly K-curve, calculate whether there is a trough in the current curve. As long as the bottom of the new cycle is not clearly and accurately found (that is, both  $b_{btc}$ and  $b_{eth}$  are false, and the initial values are both false), we will not buy cryptocurrencies until at least one bottom of one of them appears, and then we will start to buy. Based on experience, it is usually the case that Bitcoin drops to the bottom first and also rises to the peak first. When the bottom of one of the cryptocurrencies hits, we can first purchase a larger amount of that cryptocurrency, and we can also buy a small amount of the other one. The specific proportion can be set by ourselves. If both troughs occur, then use all USDT to purchase Bitcoin and Ethereum.
\item Then what we need to do is  to wait for the appearance of the peak. This period may last for two to four years. If you want to achieve greater profits, you can use traditional methods to make small portfolio adjustments during this time. We will not discuss this in detail here. Until at least one peak appears (that is,  $p_{btc}$ or $p_{eth}$  is true and the initial values are both false), then start selling positions in large quantities. After two peaks appear, we reduce positions to 0. 
\end{itemize} 
The specific algorithm details and parameter adjustments can be made according to the investor's risk tolerance.

\begin{algorithm}
	\caption{General spot trading of BTC and ETH}
	\begin{algorithmic}[1] 
		\Require $ p_{no},usdt, b_{btc}\gets \text{False}, p_{btc}\gets \text{False},b_{eth}\gets \text{False}, p_{eth}\gets \text{False}$
		\Ensure void
		\Function {Main}{$ $}
		\While{\textbf{False}==\Call{CancelAll}{$ $}}\Comment{\Call{CancelAll}{$ $} is Algorithm 1}
		\State $p_{mo}\gets$ \Call{Transform}{$p_{no}$}\Comment{real-time data conversion}
		\State \Call{Append}{$P_{btc}, P_{eth}, p_{mo}$}
		
		\If {$ \neg  (b_{btc} \land  b_{eth})$}\Comment{bottoms are not both found}
		\State $b_{btc}, b_{eth} \gets$ \Call{BottomsFound}{$P_{btc}, P_{eth}$}
		\If {$b_{btc} \land b_{eth}$}
		\State $btc, eth \gets$ \Call{BuyAll}{$usdt$}
		\ElsIf {$b_{btc} \lor b_{eth}$}
		\State $btc, eth \gets$ \Call{BuyPart}{$usdt$}
		\EndIf
		
		\Else
		\State $p_{btc}, p_{eth} \gets$ \Call{PeaksFound}{$P_{btc}, P_{eth}$}
		\If {$p_{btc} \land p_{eth}$}
		\State $usdt \gets$ \Call{SellAll}{$btc, eth$}
		\ElsIf {$p_{btc} \lor p_{eth}$}
		\State $usdt \gets$ \Call{SellPart}{$btc, eth$}
		\Else
		\State \Call {Adjust}{$usdt,btc,eth$} \Comment{traditional methods to make small portfolio adjustments}
		\EndIf
		
		\EndIf
		\EndWhile

		\EndFunction
	\end{algorithmic}
\end{algorithm}

\section{The Experimental Results on the Binance Testnet} \label{sec5}

We conducted back testing on Binance's testnet~\cite{binance} to determine that our model and algorithm parameters can accurately predict the troughs, peaks, and timing of buying and selling in the next major cycle of Bitcoin and Ethereum. To eliminate data errors, we comprehensively referred to the historical data of Yahoo~\cite{yahooFinance}, BITBO~\cite{bitboPrice} and CoinMarketCap~\cite{coinmarketcapCurrencies}, and selected the data of the two major cycles of Bitcoin between 2013 and 2018 as the main reference basis. The values of the parameters and the calculation results are shown in Table 1. The trough and peak prices are denominated in US dollars or USDT, while the trading volume is the total volume of transactions counted by Bitcoin or Ethereum. As can be seen from Table 1, our prediction results are very close to the actual values from 2019 to 2022. The actual lowest value of the new cycle basically falls at the midpoint of the prediction range, and the actual highest value is on the larger side of the prediction range. This also indicates that when the number of cryptocurrency holders increases steadily, the predictability of its overall price is very strong.

\begin{figure}[H]
	\centering
	\includegraphics[width=1.0\textwidth]{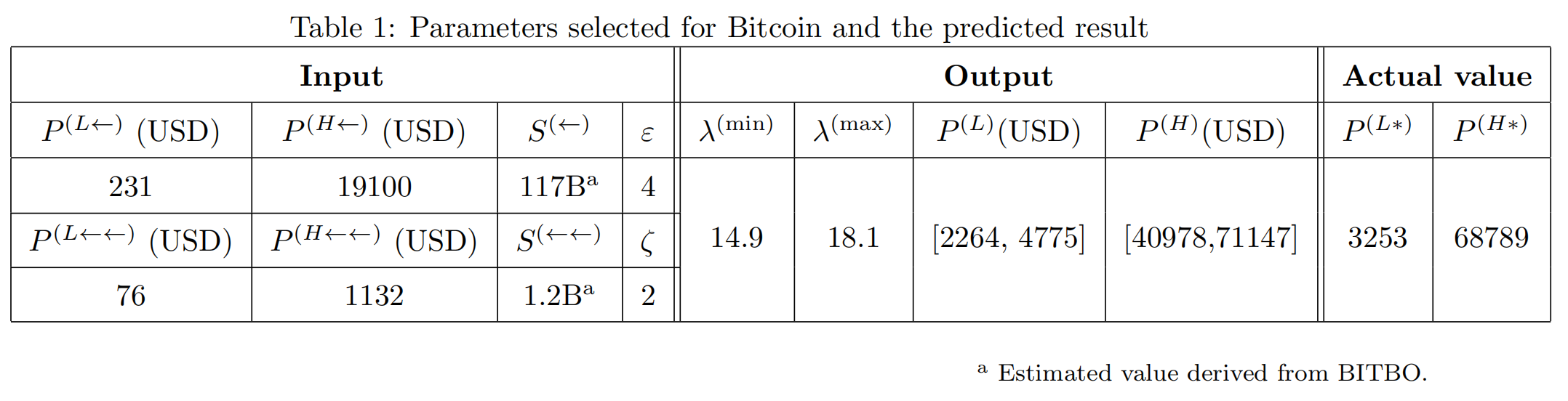}
	\label{fig:tab1}
\end{figure}

Given the late official launch of Ethereum, its price trend is unstable and cannot form three stable major cycles. Therefore, we split the major cycle of Ethereum from 2016 to 2018 into two major cycles. Also, due to the small number of its holders in the early stage causing greater volatility, we appropriately enlarged $\varepsilon$  and reduced $\zeta$. As can be seen from Table 2, our assessment of the lowest value of Ethereum's market performance from 2019 to 2022 is on the high side, but the highest value falls within our predicted range. Our algorithm underestimates its decline. This also reflects that if the number of holders of a certain cryptocurrency is insufficient, speculation will increase. Do not buy it easily.

\begin{figure}[H]
	\centering
	\includegraphics[width=1.0\textwidth]{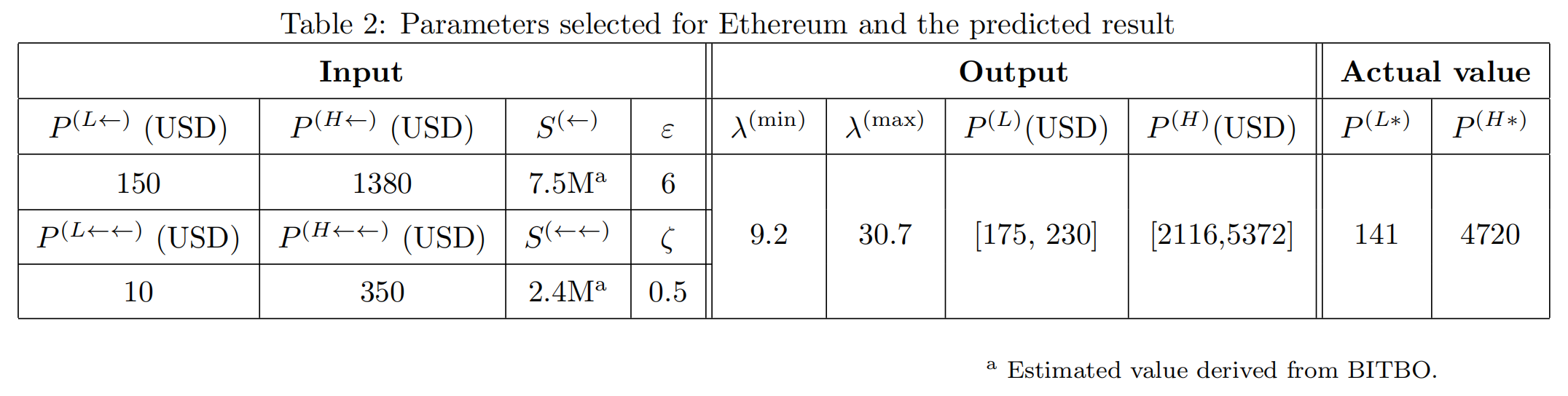}
	\label{fig:tab2}
\end{figure}

For the sake of caution, we use $70\%$ of the total USDT to purchase Bitcoin and $30\%$ to purchase Ethereum. When the algorithm gave a warning of a sharp decline in the market, we used $60\%$ USDT to purchase Bitcoin and Ethereum, and the remaining $40\%$ was used to buy cryptocurrencies at the midpoint of the trough price given by the algorithm. Similarly, we sell $60\%$ of the cryptocurrencies at the minimum value of the peak range and sell the remaining cryptocurrencies at the middle value of the peak price range. When using only this strategy, the ROI value we obtained exceeded $10$, while after conducting portfolio adjustment operations using the traditional momentum strategy, the ROI value approached $15$. The experimental results show that the prediction model based on momentum price can help investors obtain stable and safe high returns.

\begin{figure}[H]
	\centering
	\includegraphics[width=1.0\textwidth]{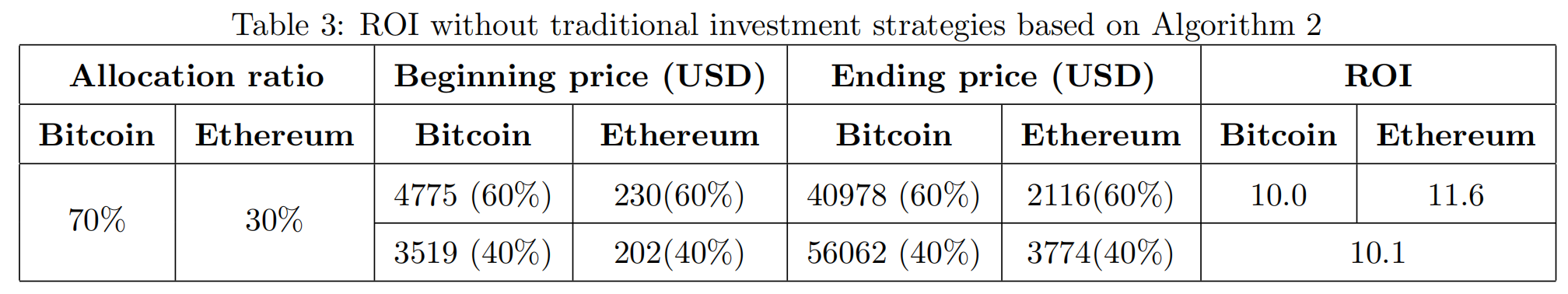}
	\label{fig:tab3}
\end{figure}

\section{Discussion} \label{sec6}

No matter how speculative the market is, as long as human nature remains unchanged, there must be a stable pattern that shows one cyclical trough and peak after another on the cryptocurrency price curve. For the first time, we have stated that the full-scale promotion of CBDC is the main factor in the decline of the cryptocurrency market. That is to say, as long as we clearly confirm the condition for withdrawing from the cryptocurrency market, do not let greed influence our short-term and long-term decisions, and aim for the maximum probability of profit rather than the goal of achieving the maximum profit, it can be said that achieving several times the profit in a major cycle is almost a $100\%$ certainty. Although our algorithm has a very high probability of making a profit, due to the significant short-term volatility of cryptocurrency prices, when using this algorithm framework, investors need to have a strong psychological endurance, such as being able to withstand a $30\%$ drop in cryptocurrency prices within a day, a $50\%$ drop within a week, or a $100\%$ drop within a month. When investors no longer view financial markets such as cryptocurrencies as casinos, but actively analyze the behavioral patterns of numerous retail investors and market manipulators, and deviate from the collective decisions of many retail investors, then it can be said that everyone's trump cards have been revealed by you. All we need to do is to optimize the details and parameters of the algorithm.

We have provided the definitions of momentum opening and closing price. However, momentum closing price is determined by the adjacent momentum opening time. Whether a more detailed definition of momentum closing price can be provided to facilitate the judgment of price trends is a question that can be further explored. When defining weighted momentum, we use a quadratic function, which performs well on Bitcoin and Ethereum. However, for other more speculative currencies, its universality needs further verification. In this paper, we only roughly provide the basis for judging the troughs and peaks of the major cycle of cryptocurrencies. More practical data is still needed to optimize the parameters. Overall, parameter tuning within a framework that accurately maps the collective behavior patterns of the cryptocurrency market can ensure continuous and stable profits.

\bibliography{refs}

\end{document}